# Astronomical Quantum-chemical Origin of Ubiquitously Observed Interstellar Infrared Spectrum due to Polycyclic Aromatic Hydrocarbon

Norio Ota

Graduate School of Pure and Applied Sciences, University of Tsukuba,
1-1-1    Tenoudai Tsukuba-city 305-8571, Japan;    n-otajitaku@nifty.com

   Interstellar infrared observation shows featured spectrum due to polycyclic aromatic hydrocarbon (PAH) at wavelength of 3.3, 6.2, 7.6, 7.8, 8.6, and 11.3 micrometer, which are ubiquitously observed in many astronomical dust clouds and galaxies. Our previous first principles calculation revealed that void induced coronene $(C_{23}H_{12})^{2+}$ and circumcoronene $(C_{53}H_{18})^{1+}$ could reproduce such spectrum very well. In this study, quantum-chemical origin was studied through atomic configuration change and atomic vibration mode analysis. By a high-speed particle attack in interstellar space, carbon void would be introduced in PAH.  Molecular configuration was deformed by the Jahn-Teller quantum effect. Carbon SP3 local bond was created among SP2 graphene like carbon network. Also, carbon tetrahedron local structure was created. Such peculiar molecular structure would be the quantum-mechanical origin. Those metamorphosed molecules would be photo-ionized by the central star's strong photon irradiation resulting cation molecules. Atomic vibration mode of cation molecule $(C_{23}H_{12})^{2+}$ was compared with that of neutral one $(C_{23}H_{12})$. At 3.3 micrometer, both molecules show C-H stretching mode and gives fairly large infrared intensity. At 6.2, 7.6, 7.8, and 8.6 micrometer bands, cation molecule show complex C-C stretching and shrinking mixing modes and remain large infrared emission. On the other hand, neutral molecule gives harmonic motion to bring cancelled small infrared emission. At 11.3 micrometer, both neutral and cation molecules show C-H bending vibration perpendicular to a molecular plane, which contributes to strong emission. Actual observed spectrum would be a sum of such different molecular size and different ionized state spectra depend on mother dust cloud and central star astronomical history.

Key words:  interstellar infrared spectrum, PAH, quantum-chemistry, coronene, circumcoronene

1, INTRODUCTION

Interstellar infrared observation shows featured spectrum due to polycyclic aromatic hydrocarbon (PAH) at wavelength of 3.3, 6.2, 7.6, 7.8, 8.6, and 11.3 micrometer, which are ubiquitously observed in many astronomical dust clouds and galaxies (Boersma et al. 2013, 2014, Sakon 2007, Leja 2016). However, any single PAH molecule or related species showing universal infrared spectrum had not yet been identified until 2014. It was found for the first time in calculation that coronene related molecule $(C_{23}H_{12})^{2+}$ suggests very similar infrared spectrum with observed one (Ota 2014, 2015a, 2017a). Based on observation of nebula NGC 2023 (Peeters 2017), we could find out that among 16 observed bands, 14 bands were successfully reproduced both for wavelength and intensity ratio (Ota 2017b). Very recently, it was found again in larger molecule that circumcoronene related PAH $(C_{53}H_{18})^{1+}$ could reproduce such ubiquitous spectrum (Ota 2018a).
  In this study, we like to focus on studying quantum-chemical mechanism and origin of such spectrum. It will be explained that carbon single void in PAH may bring serious molecular configuration change by the Jan-Teller effect. Also, molecular vibrational mode and intensity difference will be brought by molecular charge state by photo-ionization due to the central star's strong photo-illumination.

## 2, MODEL MOLECULES AND CALCULATION STEP

Initial model molecules are coronene ($C_{24}H_{12}$) and circumcoronene ($C_{54}H_{18}$) as illustrated in Figure 1. As discussed in previous papers (Ota 2018a, 2018b), these molecules would be attacked by high energy (high speed) interstellar particles, mostly proton $H^+$, which may bring a carbon void. In case of ($C_{24}H_{12}$), there are two different void positions as noted by position c and d to be ($C_{23}H_{12}$-a) and ($C_{23}H_{12}$-b), also in ($C_{54}H_{18}$) four void positions noted by a, b, c, and d to be ($C_{53}H_{18}$-a), ($C_{53}H_{18}$-b), ($C_{53}H_{18}$-c), and ($C_{53}H_{18}$-d). Such void induced PAH would be irradiated by high energy photon (over 4eV) from the new born and/or active central star and photo-ionized to be cation as ($C_{23}H_{12}$-c,-d)$^{n+}$ and ($C_{53}H_{18}$-a,-b,-c,-d)$^{n+}$. These void-induced and photo-ionized molecules should be analyzed by the first principles quantum-chemistry calculation to obtain their quantum-mechanical basic characteristics and to find out quantum origin of ubiquitously observed interstellar infrared spectrum.

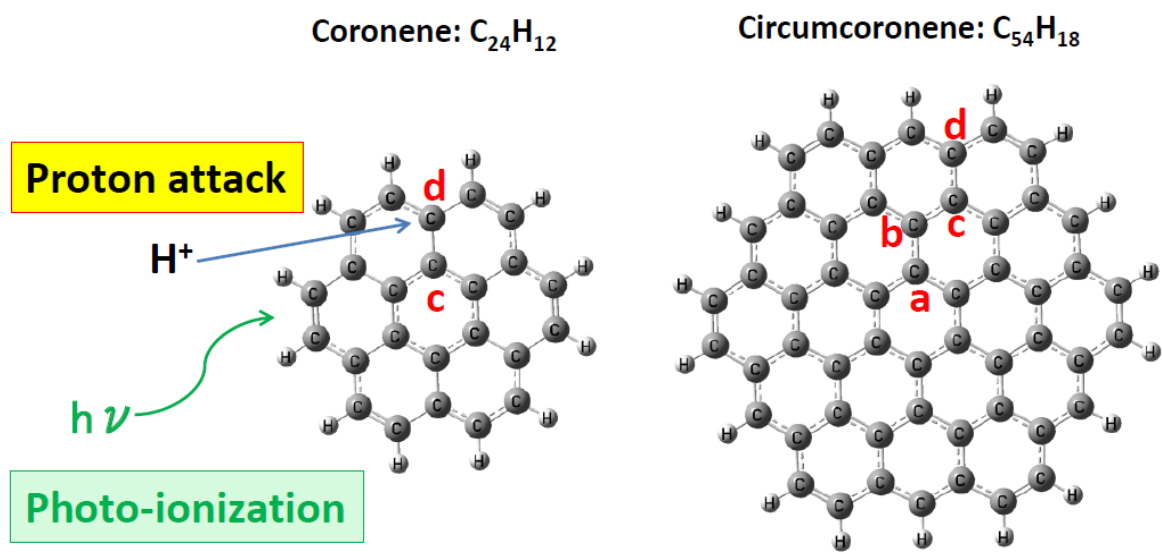

Figure-1, Initial model molecules of coronene ($C_{24}H_{12}$) and circumcoronene ($C_{54}H_{18}$) High speed interstellar proton $H^+$ may attack carbon atom and make a void at position a, b, c, and d. High energy photon would be irradiated and ionized these molecules.

## 3, CALCULATION METHOD

In quantum chemistry calculation, we have to obtain total energy, optimized atom configuration, and infrared vibrational mode frequency and strength depend on a given initial atomic configuration, charge and spin state Sz. Density functional theory (DFT) with unrestricted B3LYP functional was applied utilizing Gaussian09 package (Frisch et al. 1984, 2009) employing an atomic orbital 6-31G basis set. The first step calculation is to obtain the self-consistent energy, optimized atomic configuration and spin density. Required convergence on the root mean square density matrix was less than $10^{-8}$ within 300 cycles. Based on such optimized results, harmonic vibrational frequency and strength was calculated. Vibration strength is obtained as molar absorption coefficient ε (km/mol.). Comparing DFT harmonic wavenumber $N_{DFT}$ (cm$^{-1}$) with experimental data, a single scaling factor 0.965 was used (Ota 2015b). Concerning a redshift for the anharmonic correction, in this paper we did not apply any correction to avoid over-estimation in a wide wavelength representation from 2 to 30 micrometer.

Corrected wave number N is obtained simply by N (cm$^{-1}$) = $N_{DFT}$ (cm$^{-1}$) x 0.965.

Wavelength λ is obtained by λ (micrometer) = 10000/N(cm$^{-1}$).

Reproduced IR spectrum was illustrated in a figure by a decomposed Gaussian profile with full
width at half maximum FWHM=7cm$^{-1}$.

## 4, CALCULATED RESULTS

Molecular configuration and infrared spectrum are calculated for void induced PAH's as like $(C_{23}H_{12}\text{-c,-d})^{n+}$ (n=0, +1, +2) and $(C_{53}H_{18}\text{-a,-b,-c,-d})^{n+}$.

(1) There occurs molecular Jahn-Teller quantum deformation (Jahn and Teller 1937) in every void induced molecules. Initial void PAH has a simple void hole, but stabilized one show very complex configuration in order to minimize molecular energy. Such quantum metamorphose were illustrated in Appendix-1 to -6.
(2) Based on above stable configuration, atomic vibration induced infrared spectrum was calculated. Results are figured again in Appendix-1 to -6.
(3) Calculated infrared spectrum was compared with astronomically observed one in order to identify featured bands. In Table-1, compared result was summarized. Here, "good" means good theoretical reproduction both in wavelength and intensity ratio, also "fair" and "no good" are judged.

Based on this table, we could notice that,
(a) For every charge neutral (n=0) molecule, there are two major spectrum bands at 11.0~11.8 micrometer and 3.2~3.3 micrometer.
(b) Cationic condition is necessary to have "good" and "fair" results.

Table-1, Identification with astronomically observed infrared spectrum and calculated one.
Raw data and figures are listed in Appendix-1 to -6.

|  | Charge n=0 | n=+1 | n=+2 |
|---|---|---|---|
| $(C_{23}H_{12}\text{-c})^{n+}$ | Two peaks at 3.3 µm, 11.8 µm | Fair | Good |
| $(C_{23}H_{12}\text{-d})^{n+}$ | 3.3 µm, 11.6 µm | Fair | No good |
| $(C_{53}H_{18}\text{-a})^{n+}$ | 3.2, 11.0 | Fair | Fair |
| $(C_{53}H_{18}\text{-b})^{n+}$ | 3.2, 11.2 | Fair | No good |
| $(C_{53}H_{18}\text{-c})^{n+}$ | 3.2, 11.1 | Good | Fair |
| $(C_{53}H_{18}\text{-d})^{n+}$ | 3.2, 11.1 | Good | Fair |

## 5, QUANTUM ORIGIN-1 : CARBON SP3 BONDING

In Figure-2, calculated molecular configuration and infrared spectrum on void type-c molecules $(C_{23}H_{12}\text{-c})^{2+}$ and $(C_{53}H_{18}\text{-c})^{1+}$ were compared with astronomically observed typical example, the red rectangle nebula in milky way galaxy (Mulas 2006). Identification both for wavelength and intensity ratio at 3.3, 6.2, 7.6, 7.8, 8.6, and 12.7 micrometer bands are very well in these two molecules. Only calculated 10.9 micrometer band of $(C_{53}H_{18}\text{-c})^{1+}$ is far from observed 11.3 micrometer one.

Detailed molecular structure of $(C_{53}H_{18})^{1+}$ is shown in Figure-3. Front view shows carbon two pentagons surrounded by many carbon hexagons. Side view is Y-shape like split structure. Inserted table is atom to atom distances, and bonding angles. Number 0 Carbon (marked in red) is the central atom having SP3 like single bonds with length of 0.151~0.158nm. We could understand that specific infrared spectrum comes from carbon SP3 (S, Px, Py, and Pz electrons) like bond structure on SP2 (S, Px, and Py, electrons) graphene like pai-electron network resulting specific atomic vibration and infrared spectrum to be discussed in chapter 8.

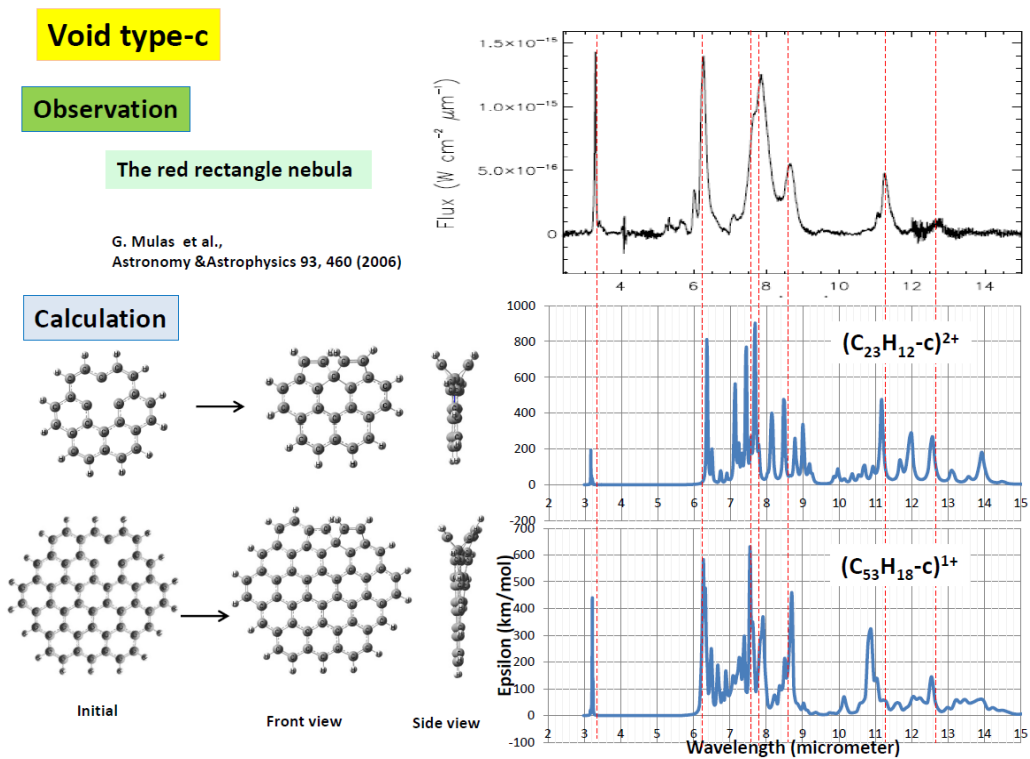

Figure-2, Void type-c molecular configuration and infrared spectrum compared with ubiquitously observed typical interstellar infrared spectrum of the red rectangle nebula.

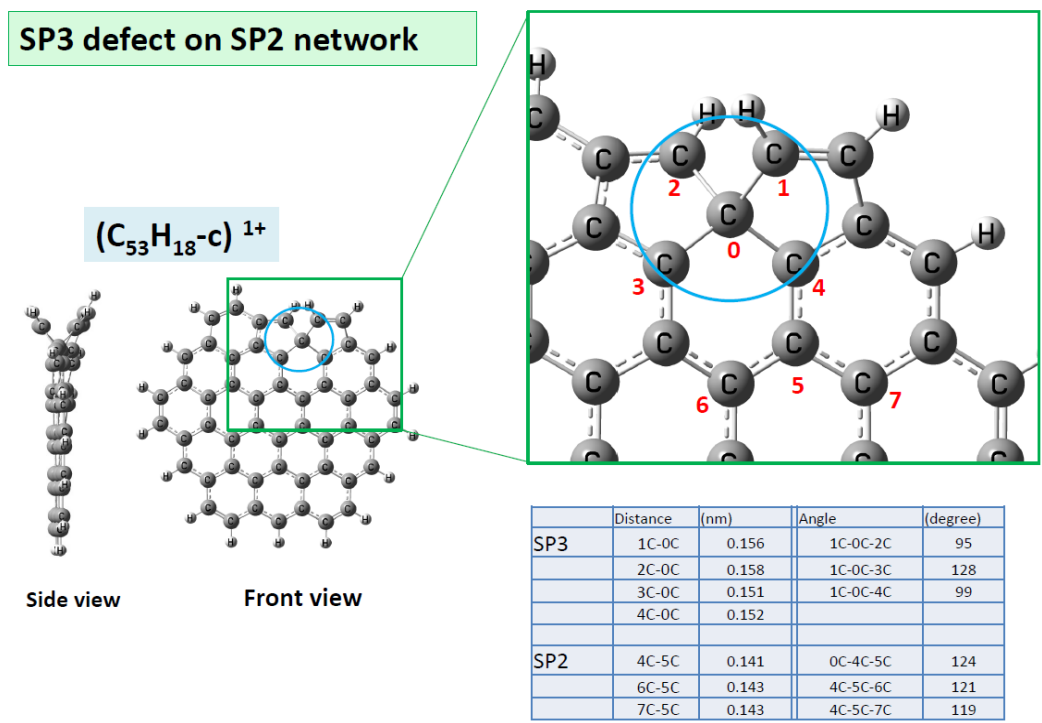

Figure-3, Molecular structure of SP3 like bonds (numbered carbon 0 to 4) among SP2 graphene like network.

## 6, QUANTUM ORIGIN-2 : CARBON TETRAHEDRON

In Figure-4, molecular configuration and infrared spectrum on void type-d molecules $(C_{23}H_{12}\text{-d})^{1+}$ and $(C_{53}H_{18}\text{-d})^{1+}$ were compared with observed spectrum of the red rectangle nebula (Mulas 2006). In case of $(C_{53}H_{18}\text{-d})^{1+}$, we could identify very well both for wavelength and intensity ratio at 3.3, 6.2, 7.6, 7.8, 8.6, and 12.7 micrometer. Only calculated 10.9 micrometer band is far from observed 11.3 micrometer. Identification on $(C_{23}H_{12}\text{-d})^{1+}$ is judged to be "fair", because of no good extra bands from 6.4 to 7.3 micrometer.

Figure-5 shows a detailed molecular structure of $(C_{53}H_{18}\text{-d})^{1+}$. Front view shows carbon one pentagon surrounded by many hexagons. It should be noted that there is extra hydrogen perpendicularly aligned to molecular plane. Side view shows slightly curved bending structure. In a zoomed figure, we can recognize carbon tetrahedron structure by a combination of carbon 0, 2, 3, and 4. It should be noted that hydrogen atom (numbered 1) plays an important role to pull up number 0 carbon, that is, to give a strained force. Inserted table is atom to atom distance, and bonding angles. We could understand that specific infrared spectrum comes from carbon SP3-H (S, Px, Py, and Pz electrons) tetrahedron structure among SP2 (S, Px, and Py electrons) graphene like network.

In 2014, Alvaro Galue et al. (Alvaro Galue 2014) proposed a theoretical model decoding astronomical infrared spectrum by introducing forced pyramid like carbon structure of modified circumcoronene. In case of cation, they could reproduce the ubiquitously observed astronomical spectrum. This is very similar with our type-d case having carbon tetrahedron structure. Difference is that our study introduced a single void bringing SP3 carbon bonds by the Jahn-Teller quantum deformation, whereas their study introduced arbitrarily forced pyramidal carbon with partially SP3 hybridized SP2 as like SP2.05.

Very recently, we opened that polycyclic pure carbon (PPC) could also contribute to reproduce 6.2, 7.6, 7.8, and 8.6 micrometer bands as shown in Figure-6 as $(C_{53}\text{-c})^{1+}$, which was obtained based on dehydrogenation from void type-c PAH $(C_{53}H_{18}\text{-c})^{1+}$ (Ota 2018a). It should be noted that even in PPC there is a carbon tetrahedron local structure, which would be brought by radical carbon strong stress. This suggests that 6.2, 7.6, 7.8, and 8.6 micrometer bands may come from pure carbon vibration. PPC molecule would be created by ultimate photo-ionization as discussed in our previous paper (Ota 2018b).

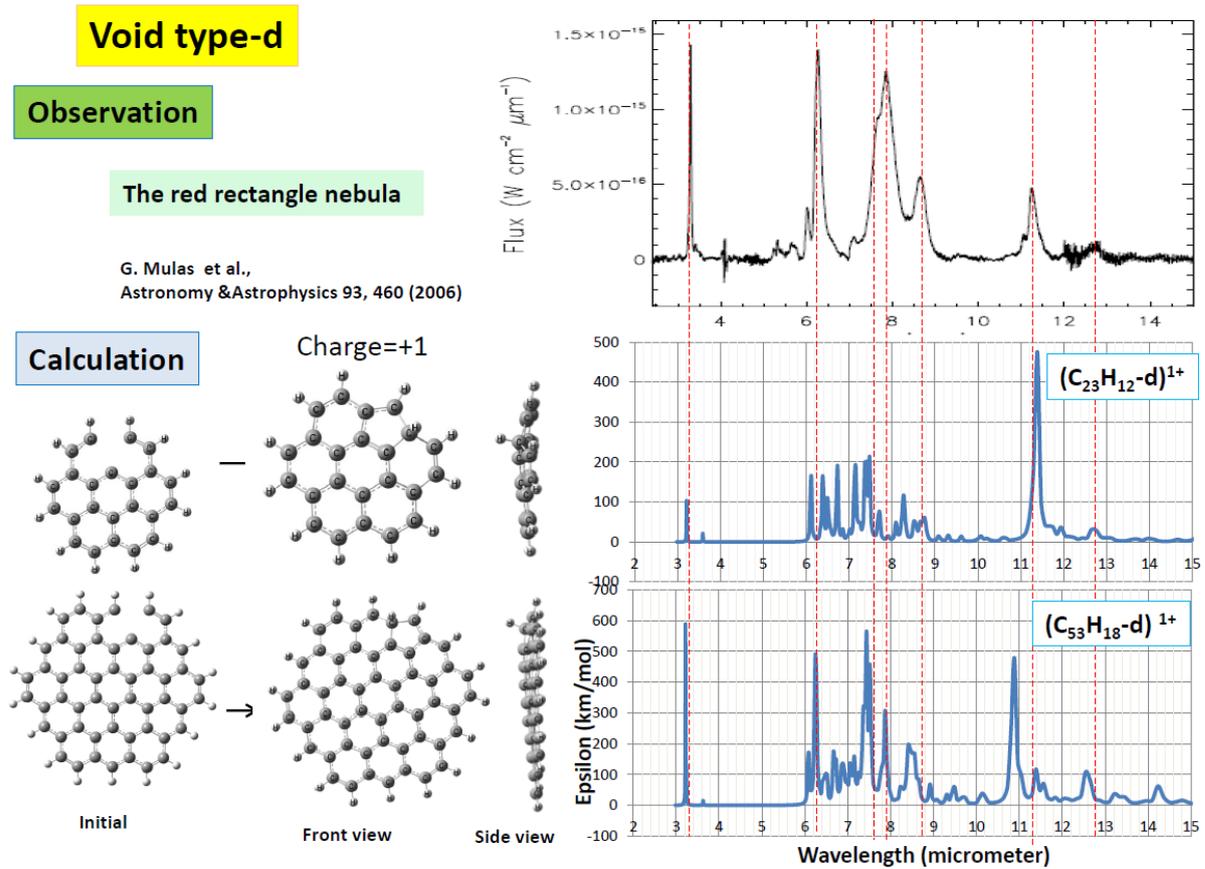

Figure-4, Void type-d molecular configuration and infrared spectrum compared with ubiquitously observed interstellar infrared spectrum, typically the red rectangle nebula.

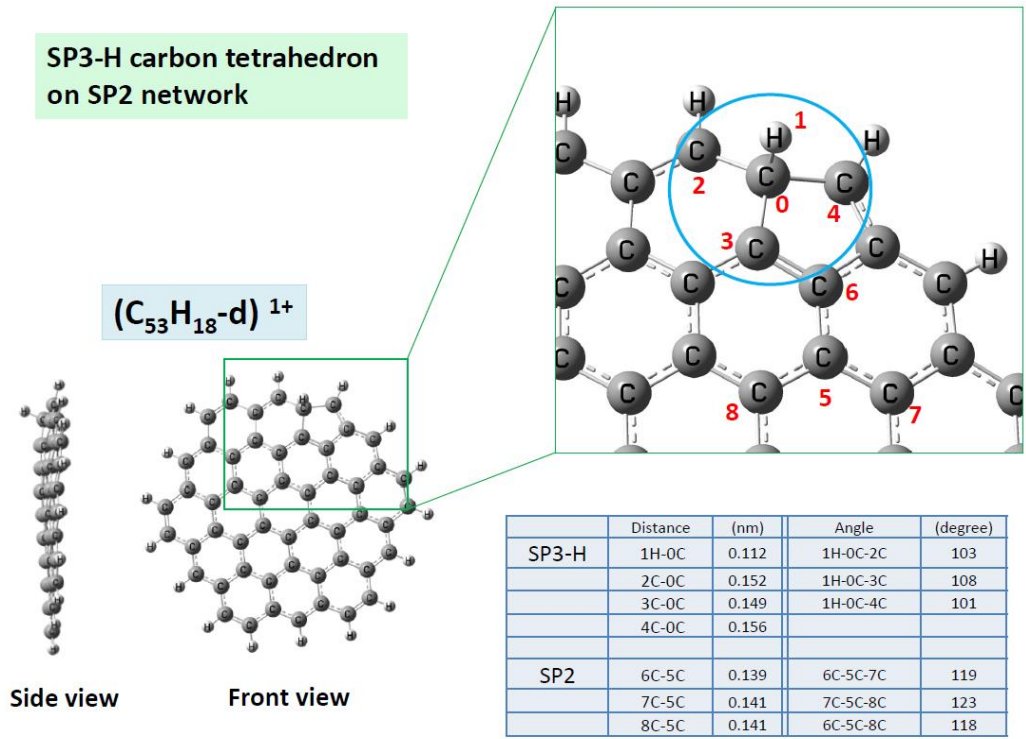

Figure-5, Molecular structure of $(C_{53}H_{18}\text{-d})^{1+}$. Blue circle shows carbon tetrahedron by SP3-H bond in SP2 graphene like network.

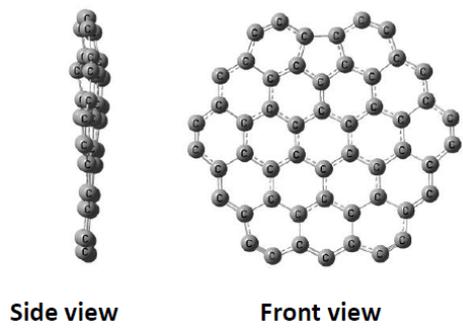

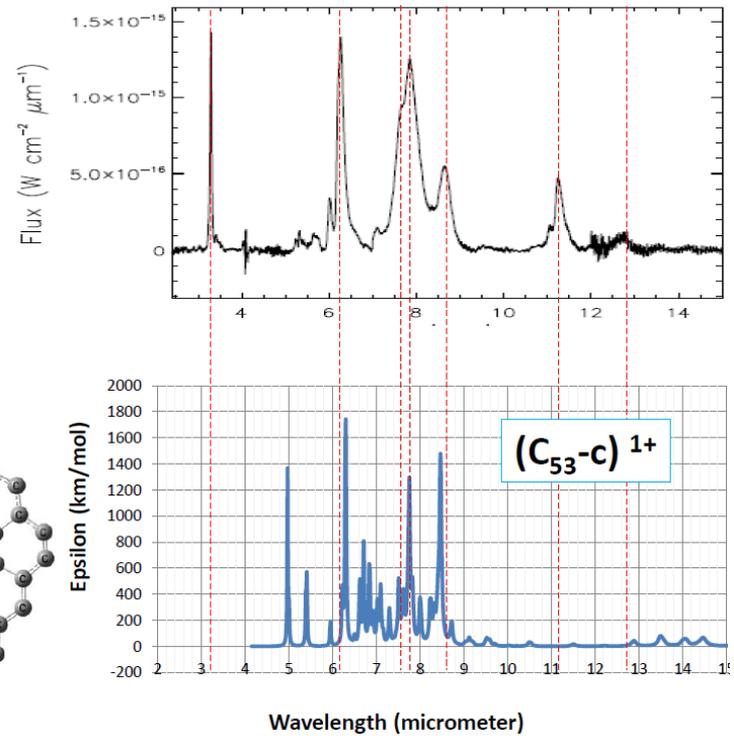

Figure-6 Polycyclic pure carbon $(C_{53}\text{-c})^{1+}$ reproducing 6.2, 7.6, 7.8, and 8.6 micrometer bands

## 7, ATOMIC CHARGE DISTRIBUTION

Spectrum difference between neutral and cation molecule is remarkable as shown in Figure-7 for $(C_{23}H_{12}\text{-c})^{0+}$ and $(C_{23}H_{12}\text{-c})^{2+}$. Atomic charge distribution and electric dipole moment are illustrated in right side. Red circle means plus charge and its size for charge amount, whereas blue circle for minus charge.
We could notice that,
(1) All of peripheral hydrogens are charged plus.
(2) All of peripheral carbons are charged minus.
(3) Biggest minus charged atom is the central carbon with SP3 bonds.
(4) Cation molecule shows larger plus charge than neutral one.
(5) Minus charge distribution of cation molecule is complex than that of neutral one.
(6) Electric dipole moment of neutral molecule is 0.65 Debye, whereas cation 1.20 Debye, vectored by blue arrow.

Above suggested differences lead to atomic vibrational mode difference and local dipole moment dynamic motion, that is, resulted to infrared spectrum difference.
For next detailed discussion, in plane coordination are shown by x and y, whereas z perpendicular to this paper.

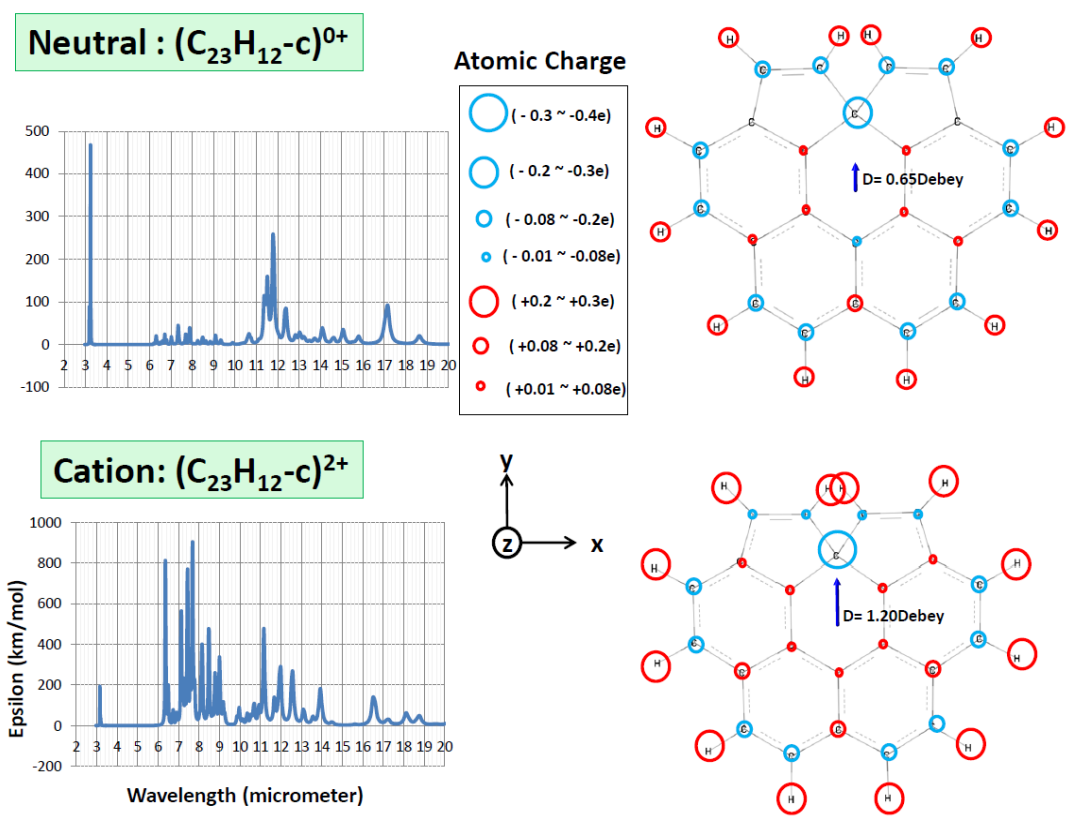

Figure-7 Charge distribution for neutral molecule $(C_{23}H_{12})^{0+}$ and cation molecule $(C_{23}H_{12})^{2+}$

## 8, VIBRATIONAL MODE COMPARISON

In order to find out detailed mechanism of featured band, it is important to compare every vibrational mode respectively between neutral molecule $(C_{23}H_{12})^{0+}$ (NE) and cation molecule $(C_{23}H_{12})^{2+}$ (CA). Every atomic vibration is illustrated in Figure-8, -9, and -10, which shows stop motion at one moment.

[1] 3.3 micrometer band (Figure-8)
Peripheral C-H shows synchronized stretching mode as symbolled by green dual-arrow. This figure shows a stop motion at one moment. All of C-H bonds are synchronously stretching at once, and shrinking at next moment. Such motion contributes strong infrared spectrum peak at $3.17\,\mu$m for NE and medium height peak at $3.15\,\mu$m for CA.

[2] 6.2 micrometer band (Figure-8)

For 6.2 micrometer band, carbon to carbon (C-C) stretching mode is major mode. In case of NE, stop motion (down left) shows synchronous stretching for symmetrically positioned carbons. Every C-C couplings are minus-minus charge, and plus-plus charge, which does not contribute on any infrared emission. Peripheral C-H shows in-plane bending associated with C-C stretching, which give harmonic canceled motion for infrared emission. Whereas, in case of CA (down right), stop motion shows complex C-C vibrations, which is a mixture of stretching (green dual arrow) and shrinking (green dual anti-arrow). Associated C-H in-plane bending does not cancelled and can contribute to x-direction emission at 6.35 micrometer.

[3] 7.6 micrometer band (Figure-9)

In case of NE at 7.33 micrometer, we can see C-C stretching and shrinking at the same time as shown in top left of Figure-9. There is no contribution from minus-minus and plus-plus charged couplings. Associated C-H in-plane bending shows symmetrical mode, which results harmonic emission cancellation. Whereas, in case of CA at 7.43 micrometer as shown in top right, there occur asymmetrical mixture of C-C stretching and shrinking, which trigger asymmetric C-H bending and contribute to large x-direction emission.

[4] 7.8 micrometer band (Figure-9)

NE molecule at 7.88 micrometer shows C-C symmetric stretching and shrinking as shown in down left of Figure-9. Associated C-H in-plane bending shows harmonic emission cancellation. In case of CA at 7.68 micrometer, there remain large y-direction contribution due to associated C-H in-plane synchronous motion at pentagon carbons .

[5] 8.6 micrometer band (Figure-10)

In case of NE at 8.49 micrometer as shown in top left of Figure-10, we can see complex C-C stretching and shrinking modes, which trigger associated C-H in-plane vibration. Total sum of those modes gives harmonic emission cancellation. Whereas in case of CA at 8.49 micrometer (top right), complex motion finally could contribute to y-direction large emission.

[6] 11.3 micrometer band (Figure-10)

Both for NE and CA, there occurs z-direction C-H synchronous bending, which contribute to large emission at 11.8 micrometer for NE and 11.2 micrometer for CA.

Figure-8, Atomic vibrational mode of 3.3 micrometer band and 6.2 micrometer band

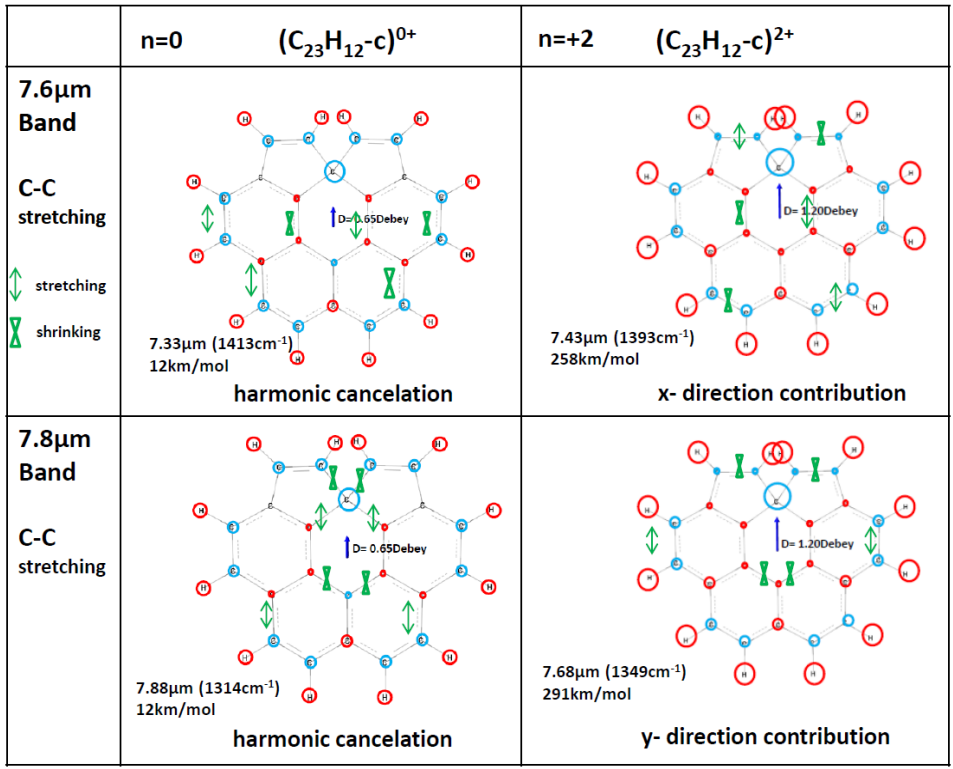

Figure-9, Atomic vibrational mode of 7.6 micrometer band and 7.8 micrometer band

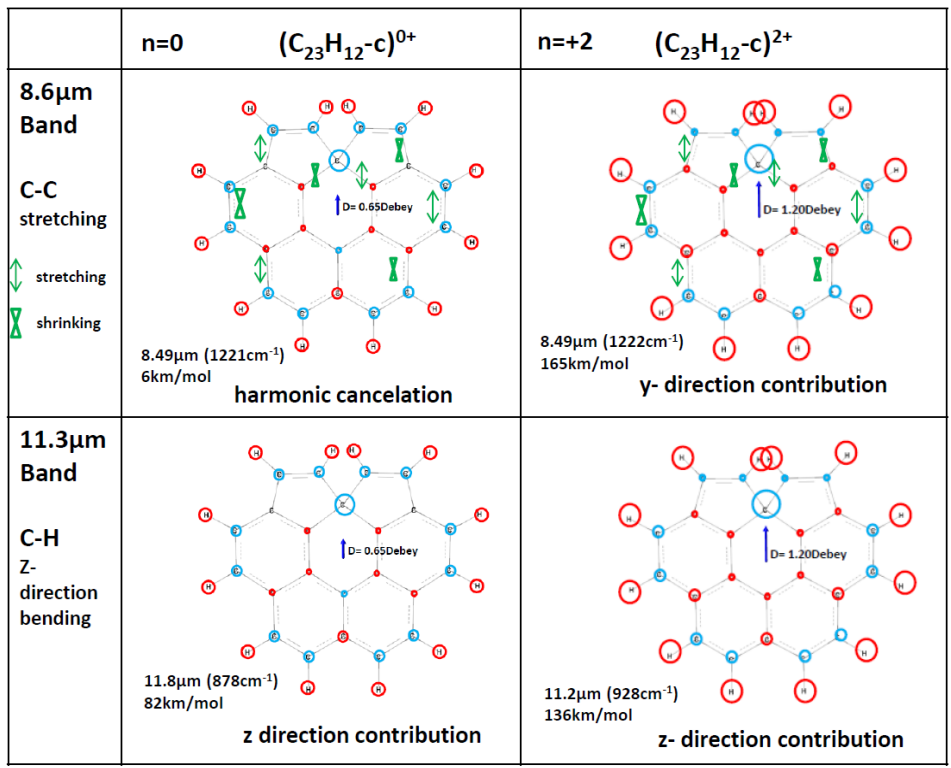

Figure-10, Atomic vibrational mode of 8.6 micrometer band and 11.3 micrometer band

## 9, CONCLUSION

Void induced cationic coronene $(C_{23}H_{12})^{2+}$ and circumcoronene $(C_{53}H_{18})^{1+}$ could reproduce ubiquitously observed interstellar infrared spectrum very well both in wavelength and intensity ratio. In this study, quantum-mechanic origin was studied through atomic configuration change and atomic vibration mode analysis.

(1) By a high speed proton attack, carbon void would be introduced in a molecule and its configuration was suddenly transformed by the Jahn-Teller quantum effect.
(2) Quantum origin-1: Void induced carbon SP3 local bond among SP2 network (graphene like carbon) brings Y-shaped 3D molecular configuration change.
(3) Quantum origin-2: Void induced hydrogen-bonded SP3-H bond causes carbon tetrahedron local structure, which brings curved bowl like molecular structure.
(4) Those metamorphosed molecule would be photo-ionized by the central star photo-illumination and became cation.
(5) Atomic vibration mode of cation molecule $(C_{23}H_{12})^{2+}$ [CA] was compared with that of neutral one $(C_{23}H_{12})$ [NE]. At 3.3 micrometer, both NE and CA show C-H stretching mode and gives fairly large infrared intensity.
(6) At 6.2, 7.6, 7.8, and 8.6 micrometer bands, CA shows complex mixture of C-C stretching and shrinking modes and remains large infrared emission. Whereas in case of NE, there occurs harmonic cancelled motion resulting little infrared emission.
(7) At 11.3 micrometer, for both in CA and NE, there occurs C-H bending motion perpendicular to a molecule plane, which contributes to strong emission.

## REFERENCES

Alvaro Galue, H., 2014, Chemical Science, 5, 2667

Boersma, C., Bregman, J.D. & Allamandola, L. J.. 2013, ApJ, 769, 117

Boersma, C., Bauschlicher, C. W., Ricca, A., et al. 2014, ApJ Supplement Series, 211:8

Bregman, B.D., et al., 2006, arXiv.org., 0604369v1

Frisch, M. J., Pople, J. A., & Binkley, J. S. 1984, J. Chem. Phys., 80, 3265

Frisch, M. J., Trucks, G.W., Schlegel, H. B., et al. 2009, Gaussian 09, Revision A.02 (Wallingford, CT: Gaussian, Inc.)

Jahn, H. and Teller, E., 1937, Proc. of the Royal Society of London, Series A, vol.161 (905), 220

Leja, J., et al., 2016, arXiv. org., 1609.09073v2

Mulas, G., et al.2006, Astronomy & Astrophysics, 93, 460

Ota, N. 2014, arXiv org., 1412.0009

Ota, N. 2015a, arXiv org., 1501.01716

Ota, N. 2015b, arXiv org., 1510.07403

Ota, N. 2017a, arXiv org., 1703.05931

Ota, N. 2017b, arXiv org., 1704.06197

Ota, N. 2018a, arXiv org., 1801.06305

Ota, N. 2018b, arXiv org., 1803.09035

Peeters, E., et al. 2017, arXiv org. 1701.06585v1

Sakon, I, et al., 2007, Publ. Astron. Soc. Japan, 59, S483

Author profile
  Norio Ota PhD, Senior Professor, University of Tsukuba, Japan,
  Material Science, Optical data storage devices

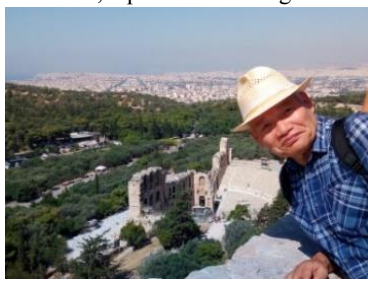

Appendix-1

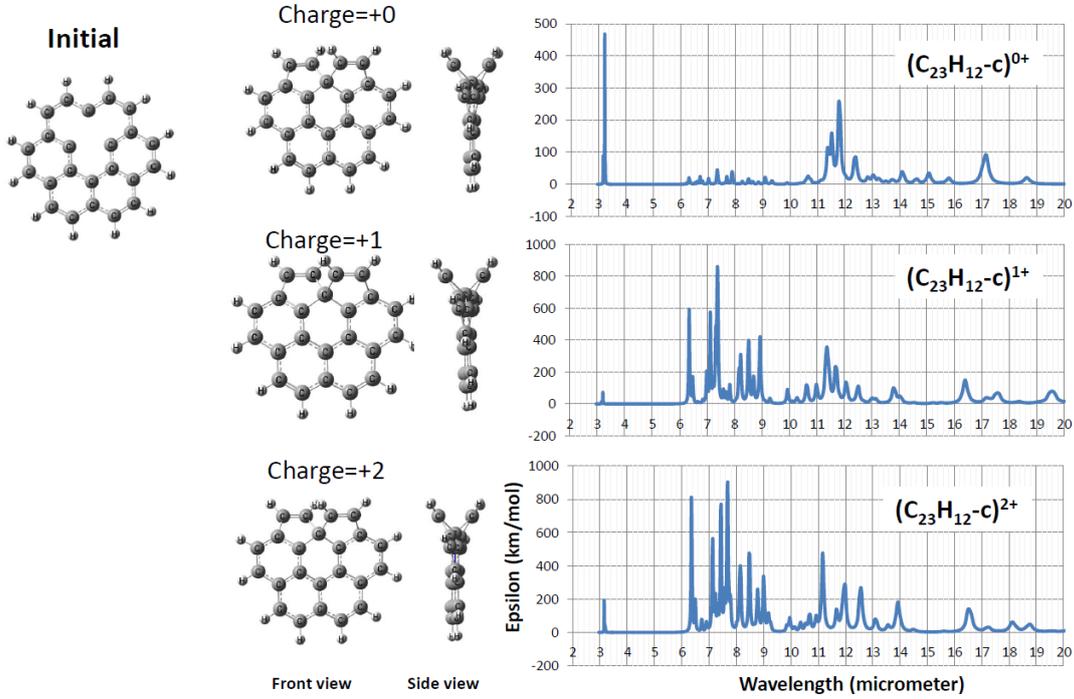

Appendix-2

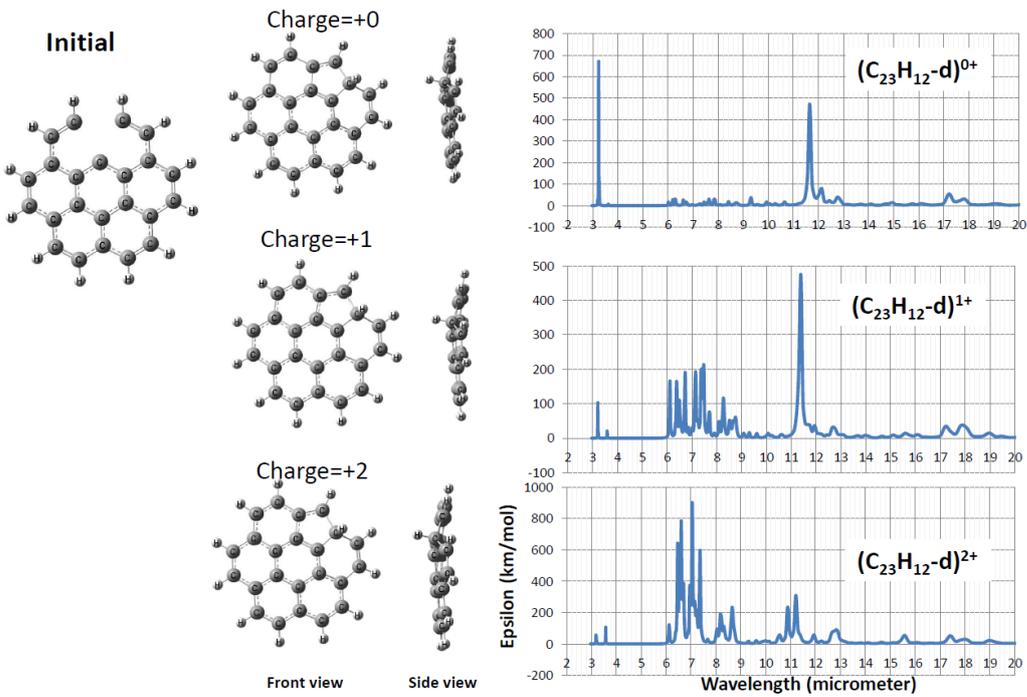

Appendix-3

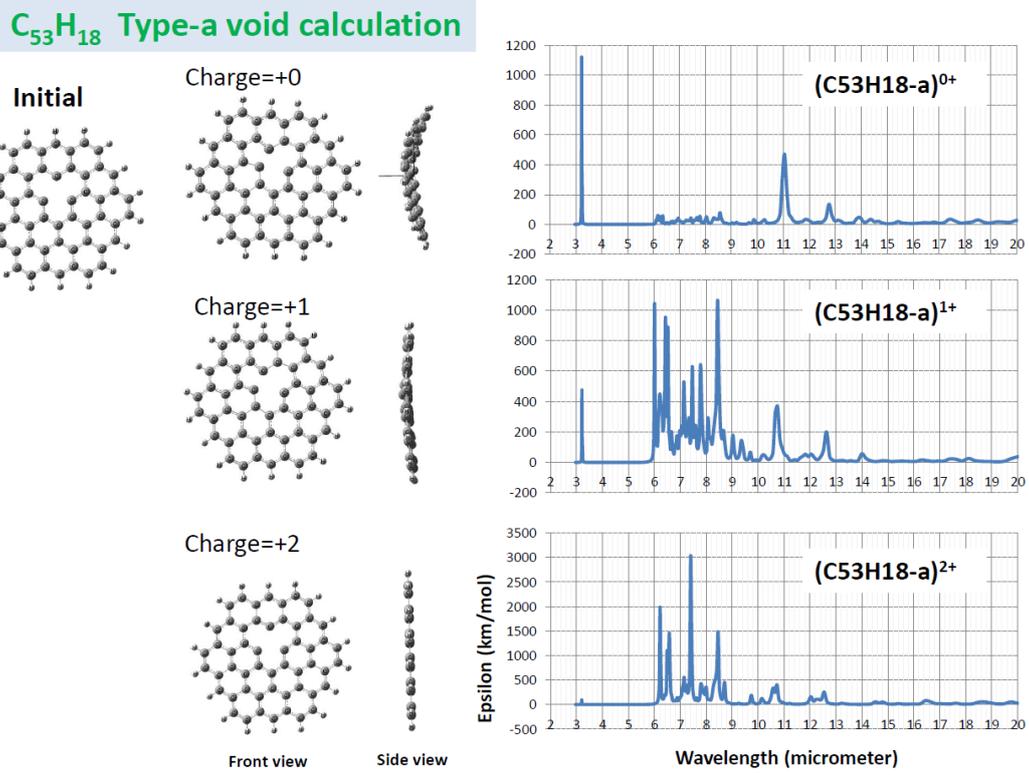

Appendix-4

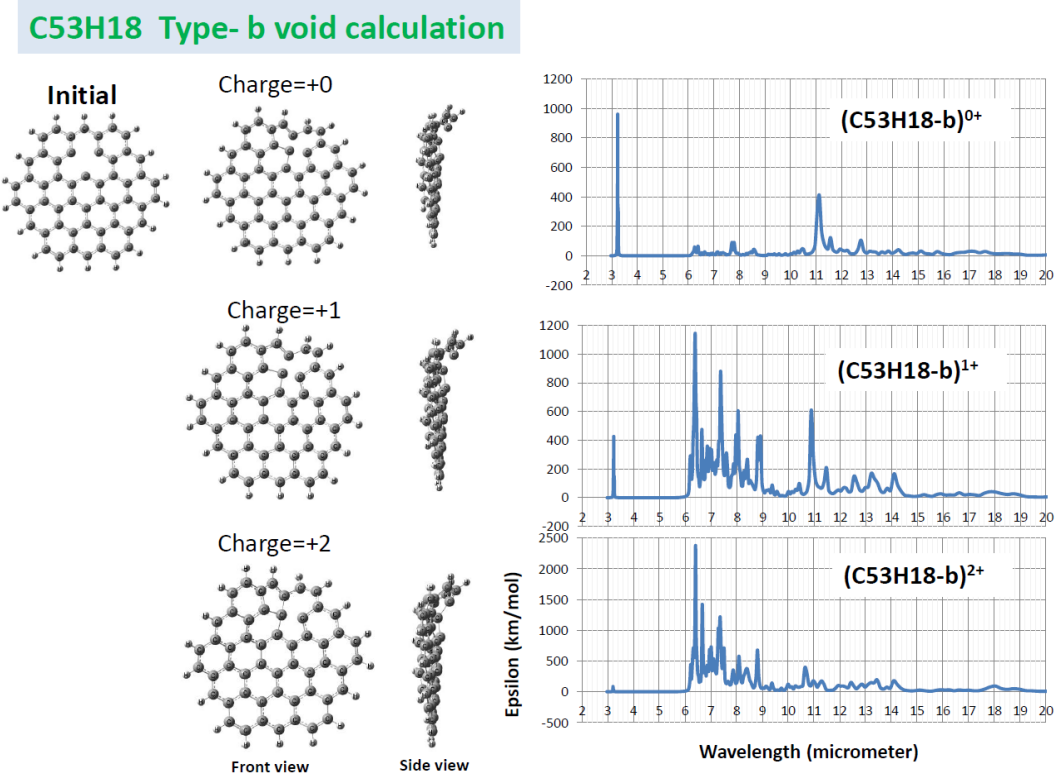

Appendix-5

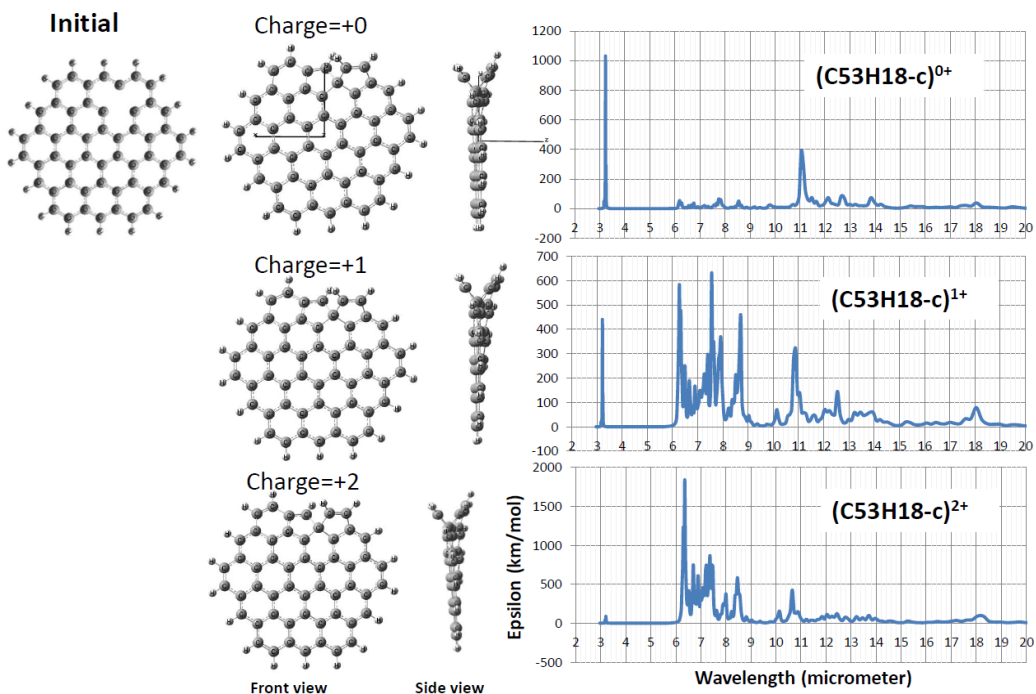

Appendix-6

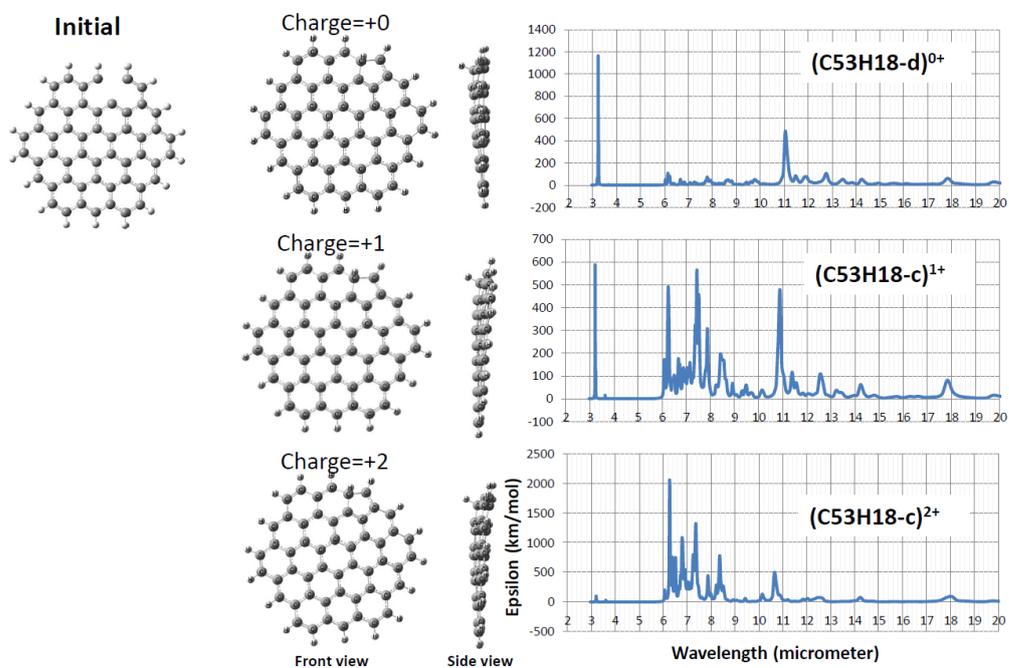

**Submitted to arXiv.org           August       , 2018 by Norio Ota**